\documentclass[useAMS]{mn2e}
\usepackage{epsfig}

	\newcommand{\chandra}{{\it Chandra}}
	\newcommand{\hst}{{\it HST}}

	\newcommand{\xmm}{{\it XMM-Newton}}
	
	\newcommand{\lum}{\thinspace\hbox{$\hbox{erg}\thinspace\hbox{s}^{-1}$}}

	\title[Localization of the X-ray Source in the Globular Cluster G1 with \chandra]{Localization of the X-ray Source in the Globular Cluster G1 with \chandra}
	\author[A.~K.~H.~Kong et al.]{A.~K.~H.~Kong,$^{1}$\thanks{E-mail: akong@phys.nthu.edu.tw}\thanks{Kenda Foundation Golden Jade Fellow} C.~O.~Heinke,$^{2}$ R. Di Stefano,$^{3}$ H.~N.~Cohn,$^{4}$ P.~M.~Lugger,$^{4}$ 
	\newauthor P. Barmby,$^{5}$  W.~H.~G.~Lewin,$^{6}$ F.~A.~Primini$^{3}$\\
	$^{1}$ Institute of Astronomy and Department of Physics, National Tsing Hua University, Hsinchu 30013, Taiwan\\
	$^{2}$ Department of Physics, University of Alberta, 11322-89 Avenue, Edmonton, AB T6G 2G7, Canada\\
	$^{3}$ Harvard-Smithsonian Centre for Astrophysics, 60 Garden Street, Cambridge, MA 02138, USA\\
	$^{4}$ Astronomy Department, Indiana University, 727 East 3rd St., Bloomington, IN 47405, USA\\
	$^{5}$ Department of Physics \& Astronomy, University of Western Ontario, London, ON N6A 3K7, Canada\\
	$^{6}$ Kavli Institute for Astrophysics and Space Research, Massachusetts Institute of Technology, Cambridge, MA 02139, USA
	}
	
\begin{document}
	
	%\date{Accepted 1988 December 15. Received 1988 December 14; in original form 1988 October 11}

	\pagerange{\pageref{firstpage}--\pageref{lastpage}} \pubyear{2010}

	\maketitle

	\label{firstpage}

\begin{abstract}
We report the most accurate X-ray position of the X-ray source in the giant globular cluster G1 in M31 by using the \chandra\ X-ray Observatory, {\it Hubble Space Telescope (HST)}, and Canada-France-Hawaii Telescope (CFHT). G1 is clearly detected with \chandra\ and by cross-registering with \hst\ and CFHT images, we derive a $1\sigma$ error radius of $0\farcs15$, significantly smaller than the previous measurement by \xmm. We conclude that the X-ray emission of G1 is likely to come from within the core radius of the cluster.
We have considered a number of possibilities for the origin of the X-ray emission but can rule all but two scenarios out: 
it could be due to either accretion onto a central intermediate-mass black hole (IMBH), or an ordinary low-mass X-ray binary (LMXB). Based on the X-ray luminosity and the Bondi accretion rate, an IMBH accreting from the cluster gas seems unlikely and we suggest that the X-rays are due to accretion from a companion. Alternatively, the probability that a $1.5 M_\odot$ cluster LMXB lies within the 95 per cent X-ray error circle is about 0.7. Therefore we cannot rule out a single LMXB as the origin of the X-ray emission. While we cannot distinguish between different models with current observations, future high-resolution and high-sensitivity radio imaging observations will reveal whether there is an IMBH at the centre of G1.

\end{abstract}


\section{Introduction}
Intermediate-mass black holes (IMBHs) have been a subject of debate for a long time. If they exist, IMBHs represent the long sought after link between
stellar-mass and super-massive black holes. Until recently, we only have had indirect evidence for IMBHs via X-ray observations. For instance, ultraluminous ($L_X>10^{39}$\lum) or hyperluminous ($L_X>10^{41}$\lum) X-ray sources have been the best candidates (e.g. Wolter et al. 2006;  Farrell et al. 2009) based on 
their X-ray luminosities which, assuming isotropic emission, are
far in excess of the Eddington limit for stellar mass black holes. Furthermore, X-ray spectroscopy and X-ray timing also provide some support for IMBHs (e.g. Miller et al. 2004; Kong \& Di Stefano 2005; Strohmayer \& Mushotzky 2009). However, we still require dynamical evidence in order to confirm solidly that IMBHs exist.

The only dynamical mass measurement claimed for an IMBH candidate is the globular cluster G1 in M31. G1 is the most luminous star cluster in the Local Group, and also one of the most massive at $(4–-7)\times 10^6\,M_\odot$ (Ma et al. 2009; Barmby et al. 2007). Based on Keck and {\it Hubble Space Telescope (HST)} observations, it has been claimed that G1 hosts a $\sim 2\times10^4 \, M_\odot$ object at the core (Gebhardt et al. 2002,2005). However, the suggestion of an IMBH is controversial and has been challenged by 
Baumgardt et al. (2003). Recently, X-ray emission near the core of G1 has been discovered based on \xmm\ observations (Trudolyubov \& Priedhorsky 2004; Pooley 
\& Rappaport 2006; Kong 2007) and it is suggested that the X-rays come from Bondi accretion from cluster gas onto a central IMBH. However, it is also possible that the X-ray emission is due to an ordinary low-mass X-ray binary (LMXB), or a collection of faint LMXBs near the core. By refining the relative astrometry between \xmm\ and \hst\ data, Kong (2007) concluded that we cannot distinguish between these two scenarios. As we will show in this Letter, if there exists an IMBH in G1, accretion likely comes from a companion star (see, e.g. Patruno et a. 2006 for a scenario with a giant).

Alternatively, radio observations may be able to provide additional information about the nature of the X-ray source in G1. Ulvestad et al. (2007) employed the Very Large Array (VLA) to obtain a low-resolution ($\sim 3''$ beam size at 8.4 GHz) image of G1 and a radio source was detected near the \xmm\ source, about an arcsecond from the cluster core. The radio/X-ray flux ratio of G1 ($\sim 5\times10^{-5}$) is at least a few hundred times higher than that expected for a LMXB near the cluster centre ($\sim 5\times10^{-8}$; see Fender \& Kuulkers 2001), but is consistent with the expected value for accretion onto a $2\times10^4 M_\odot$ IMBH (Merloni et al. 2003). Furthermore, the radio/X-ray flux ratio is much lower than that of supernova remnants ($\sim 10^{-2}$), but it is consistent with a pulsar wind nebula (Ulvestad et al. 2007).

However, it is still a question whether the X-ray and radio emission come from the same source because the resolution of both observations do not have sufficient accuracy to determine the precise position. While the VLA is still in its short baseline configuration and MERLIN is being upgraded, the first step is to use \chandra\ to obtain an accurate position for
the X-ray source.

In this Letter, we localized the X-ray emission of G1 by performing precise relative 
astrometry using \chandra, \hst, and Canada-France-Hawaii Telescope (CFHT). In Section 2, we describe our X-ray and optical observations and data analysis. We present the localization of the X-ray source in Section 3. We finally discuss the nature of the X-ray source in Section 4.

\section{Observations and Data Analysis}
\subsection{\chandra}
We observed G1 with the {\it Chandra X-ray Observatory} on 2008 September 30 for a total exposure time of 35 ks (ObsID 9525). The observation was taken using the Advanced CCD Imaging Spectrometer array (ACIS-S) with the telescope aim point at G1. Data were telemetered in the very faint mode and were collected with a frame transfer time of 3.2 s. We used CIAO version 4.1\footnote{http://cxc.harvard.edu/ciao/}, ACIS Extract\footnote{http://www.astro.psu.edu/xray/docs/TARA/ae\_users\_guide.html}, and XSPEC version 12.5\footnote{http://heasarc.gsfc.nasa.gov/docs/xanadu/xspec/}  packages to perform data reduction and analysis. We reprocessed the raw data to make use of the very faint mode. In order to reduce the background, only events with photon energies in the range of 0.3–-7.0 keV were included in our analysis. We also inspected the background count rates from the S1 chip and no flaring event was found in the data set. In this paper, we only consider the S3 chip of ACIS-S.

Discrete sources in the Chandra images were found with {\it wavdetect} (Freeman et al. 2002) together with exposure maps. We performed source detection on the 0.3–-7 keV image. We set the detection threshold to be $10^{-6}$, corresponding to less than one false detection due to statistical fluctuations in the background. We performed source detection using sequences of wavelet scales that increased by a factor of $\sqrt{2}$ from scales 1 to 16. A total of 28 X-ray sources were detected. The X-ray source in G1 was clearly detected with 126 counts.

We extracted the energy spectrum from a $2''$ circular region centred on G1. For the background, we selected a source-free region with a radius of $15''$. Response matrices were generated by CIAO. 
We then fitted the background-subtracted spectrum with an absorbed power-law model. In order to employ $\chi^2$ statistics, the spectrum was grouped into at least 10 counts per spectral bin. 
Since the statistics in the spectrum are poor, and therefore it can be fit equally well with numerous spectral models (e.g. power-law, bremsstrahlung, and thermal plasma), although a blackbody model can be ruled out. However, the absorbed power-law model is adopted as it is typical for a globular cluster X-ray source.
The power-law model provides the best fit ($\chi^2/dof=7.56/9$) to the data with $N_H=4.7^{+7.6}_{-4.7}\times10^{20}$ cm$^{-2}$ and $\Gamma=1.8^{+0.5}_{-0.4}$ (90 per cent confidence level). The 0.3--7 keV unabsorbed luminosity is $(2.3^{+0.8}_{-0.5})\times10^{36}$\lum assuming a distance of 780 kpc (Macri et al. 2001). The X-ray spectrum and luminosity are very typical for a globular cluster X-ray source in M31 (Kong et al. 2002a; Di Stefano et al. 2002). 

The X-ray luminosity as well as the spectrum of G1 are consistent with previous \xmm\ observations (Pooley \& Rappaport 2006) indicating it is a persistent source. We also calculated the Kolmogorov-Smirnov statistic to examine whether G1 is a variable source during our \chandra\ observation; there is over 25 per cent probability that G1 is a constant source.  We therefore cannot reject the null hypothesis that G1 is a constant source. We also applied the Gregory-Loredo variability algorithm (Gregory \& Loredo 1992) in CIAO to examine our G1 data and the variability index is 0, implying no variability.

\begin{figure*}
	\centering
	\psfig{file=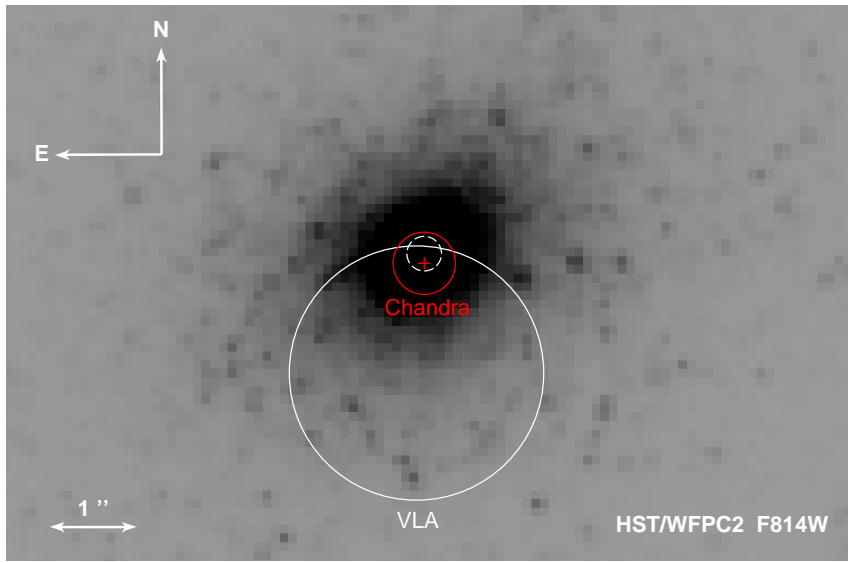,width=112mm}
	\caption{\hst\ WFPC2 F814W images of G1. The red circle is the 95 per cent error circle ($0\farcs36$ in radius) of the \chandra\ source and its centre is marked by a plus sign. The white dash circle at the centre is the core radius ($0\farcs2$; Ma et al. 1997; Barmby et al. 1997) of G1, while the white circle is the 95 per cent error circle ($1\farcs47$ in radius) of the VLA source.}
\end{figure*}

\subsection{Optical Observations}

G1 was observed with the \hst\ Wide Field Planetary Camera 2 (WFPC2) on 1995 October 2 with a total integration time of 30 minutes in the F814W filter. We downloaded the F814W image from the Hubble Legacy Archive\footnote{http://hla.stsci.edu/} for which cosmic-ray free, science-quality images are dithered, co-added, and corrected for astrometry. By comparing to the 2MASS catalogue (Skrutskie et al. 2006), the astrometry of the image is better than $0\farcs2$. Although G1 was also observed with the Advanced Camera for Surveys (ACS) in High Resolution Channel (HRC) mode (see Kong 2007), we do not use the ACS/HRC data because they have a much smaller field-of-view ($29''\times26''$) which is not useful for refining the relative astrometry with ground-based telescope and \chandra\ data.

Since the field-of-view of WFPC2 is small compared to \chandra, it is difficult to improve the relative astrometry between \hst\ and \chandra. We therefore obtained a wide-field optical image centred on G1 with the MegaPrime/MegaCam at the CFHT on 2008 September 5. The MegaCam has an array of 36 CCDs, giving a total of 1 degree by 1 degree field-of-view. We obtained a series of $i'$-band images under a seeing condition of $0\farcs7$ with a total exposure time of 721 seconds. The images were processed by Terapix\footnote{http://terapix.iap.fr/} which provides coadded and astrometrically calibrated images for data analysis. By matching over 3000 stars in the field with the 2MASS catalog, the astrometry of the CFHT image is accurate to about $0\farcs3$.

\section{Localization of the X-ray Source in G1}

To localize the precise position of the X-ray source in G1, we have to improve the relative astrometry of the X-ray and optical images. To achieve this, we need to match the \chandra, \hst, and CFHT images into a common reference frame (see, e.g. Lu et al. 2009). We first compared the \chandra\ image with the CFHT image. 
By comparing the Chandra source list with the CFHT image, we found 5
likely matches (in addition to G1) that are optically bright and
isolated.
They are likely to be foreground stars or background galaxies. Based on these 5 matches, we corrected the astrometry of the \chandra\ image using the IRAF\footnote{http://iraf.noao.edu/} task {\it ccmap}. The resulting registration errors are $0\farcs062$ in R.A. and $0\farcs117$ in declination.

We next transformed the astrometry of WFPC2 to the CFHT image. We found 55 stars in both images that appeared stellar and unblended. We obtained an astrometric solution giving residuals of $0\farcs037$ in R.A. and $0\farcs036$ in decl. After registering all the images to the CFHT image, we located the optical centre of G1 in the WFPC2 images and the position of the X-ray source. We determined the centroid of G1 in the WFPC2 image (R.A.=00h32m46.532s, 
decl.=+39d34m40.58s with $1\sigma$ error of $0\farcs002$) by fitting the elliptical isophotes to the globular cluster using the IRAF task {\it ellipse}. 

For the X-ray position, we used ACIS Extract to derive the mean position of events within the 90 per cent point-spread-function in the 0.3--7 keV band, yielding R.A.=00h32m46.532s, decl.=+39d34m40.47s with $1\sigma$ statistical error radius of $0\farcs04$. We then determined the $1\sigma$ radius 
error circle ($0\farcs15$) of the \chandra\ position of G1 by computing the 
quadratic sum of the positional uncertainty for the X-ray source ($0\farcs04$), the 
registration error between WFPC2 and CFHT images ($0\farcs05$), and the residuals between \chandra\ and CFHT alignment ($0\farcs13$). Hence, the 95 per cent radius (2-dimensional) error circle will be $0\farcs36$. Figure 1 shows the WFPC2 images 
of G1 and the 95 per cent radius X-ray error circle. 

\begin{table*}
	\centering
	\footnotesize
\caption{Comparison of G1 (assuming X-ray emission from G1 is from the putative IMBH), supermassive black holes, and stellar-mass black holes}
\begin{tabular}{cccccccc}
	\hline
	Source & $M_{BH} (M_\odot)$ & log $L_X$ & log $(L_X/L_{Edd})$ & log $(L_B/L_{Edd})$ & log $(L_X/L_{B})$ & log $L_R/(L_X/L_{Edd})$ & References\\
	   (1)    &  (2)  & (3) & (4) & (5) & (6) & (7)\\
	\hline
 	G1 & $2\times10^4$? & 36.3 & -6.11 & -4.11&-2.00 &38.6 &This paper\\
	Sgr A* & $2.6\times10^6 $& 33.3 -- 35 & -11.22 -- -9.52 & -3.85 & -7.26 -- -5.57& 42 -- 43.7& 1, 2\\
	M31* & $1.4\times10^8 $& 35.8 -- 37.3 & -10.52 -- -9.00 & -4.36&-5.45 -- -3.93 & 41.3 -- 42.8 & 3\\
	M32* & $2.5\times10^6 $& 36 & -8.52 & -5.93 & -2.83 & $< 41.8$ & 4\\
	V404 Cyg & 10 & 33 -- 33.9 & -6.11 -- -5.22 & ---& --- &33.8 -- 34.6& 5, 6\\
	A0620--00 & 10 & 30.5& -8.62& ---& --- &35.5 & 7, 8\\
	GS2000+25 & 7 & 30.4 & -8.44 & --- &--- & --- &9\\
	\hline
\end{tabular}
\par
\medskip
\begin{minipage}{0.95\linewidth}
	{\bf Notes.} Column 1: name of the object; Column 2: Black hole mass; Column 3: X-ray luminosity (\lum); Column 4: X-ray to the Eddington luminosity ratio; Column 5: Bondi to Eddington luminosity ratio; Column 6: X-ray to the Bondi luminosity ratio; Column 7: ratio between radio luminosity and $L_X/L_{Edd}$.\\
	{\bf References.} (1) Baganoff et al. 2001; (2) Merloni et al. 2003; (3) Garcia et al. 2010; (4) Ho et al. 2003; (5) Bradley et al. 2007; (6) Gallo et al. 2005; (7) Kong et al. 2002b; (8) Gallo et al. 2006; (9) Garcia et al. 2001\\
\end{minipage}
\end{table*}

\section{Discussion}
By utilizing \chandra, \hst/WFPC2, and CFHT/MegaCam data, we determined the precise position of the X-ray emission of G1. The X-ray source, previously seen by \xmm\ (see Pooley \& Rappaport 2006; Kong 2007), is very close ($\sim 0\farcs11$) to the cluster centre. Based on the calculation by Pooley and Rappaport (2006), if the X-ray emission is from Bondi accretion of ionized cluster gas by a central IMBH, the X-rays should come from the 
central 50 milli-arcsecond of the cluster. However, given the uncertainty of the error circle, we cannot rule out that the X-ray source is slightly offset from the cluster core (see Figure 1). If the X-ray emission is not from a central IMBH, it could come from a luminous LMXB located within or very close to the core radius ($0\farcs21$ in Ma et al. 2007; $0\farcs19$ in Barmby et al. 2007) of G1. As Kong (2007) points out, nearly half of the LMXBs in Galactic globular clusters are found within the core radius; it would not be surprising to find a luminous LMXB within the core of G1.

We estimated the probability of a $1.5 M_{\odot}$ object being found in the X-ray error circle. The mass of $1.5 M_{\odot}$ is the average quiescent LMXB mass, estimated from the radial distribution of 20 quiescent LMXBs in seven globular clusters (Heinke et al. 2003).
We adopted the single-mass
King model fit from Ma et al. (2007), which has $r_c = 0.21''$ and also
considered the core radius value of $r_c = 0.19''$ from Barmby et al.
(2007).  We assumed that the King model profile describes the
distribution of turnoff-mass objects; we took the turnoff mass to be
$0.9 M_\odot$.  This gives a mass ratio of $q=1.67$ between the LMXBs and
the turnoff-mass stars. It is straightforward to integrate the radial density profile for
the LMXBs analytically.  As a check, half of the probability is
within the core radius for $q=1.67$.  We then integrated the
probability over angle numerically, to compute the probability
within the off-centre X-ray error circle. We found that a $1.5 M_\odot$ LMXB would have a probability of 0.756 ($r_c = 0.19''$) and 0.718 ($r_c = 0.21''$) of being found within our 95 per cent confidence \chandra\ error circle.
Therefore the possibility that the X-ray source represents a single LMXB is significant.

If the X-ray source is from an accreting IMBH, the observed luminosity ($2\times10^{36}$\lum) and the spectrum would be consistent with a black hole in the hard state (Remillard \& McClintock 2006).  There are also two possibilities if G1 has an accreting IMBH. The X-ray emission could be due to accretion either from cluster gas (Pooley \& Rappaport 2006), or from a companion star. 

We list in Table 1 a few examples of nearby supermassive black holes and dynamically confirmed stellar-mass black holes for comparison. All three quiescent supermassive black holes in the local universe have very low Eddington ratios of $10^{-11} - 10^{-9}$. On the other hand, quiescent stellar-mass black holes tend to have a higher Eddington ratio. V404 Cyg is the most X-ray luminous stellar-mass black hole while A0620--00 and GS2000+25 are the faintest ones (with detection). 
The Eddington ratio of G1 (assuming a mass of $2\times10^4 M_\odot$) is at least two orders of magnitude higher than that of quiescent supermassive black holes, but is in the range of stellar-mass black holes. We also compare the  X-ray luminosity in units of Bondi luminosity. This is a better indicator of accretion luminosity than the Eddington ratio, as it relates the X-ray luminosity to the available mass transfer rate. For G1, it is about 0.01 (see Pooley \& Rappaport 2006), which is substantially higher than quiescent supermassive black holes (see Table 1). It is an indication that the accretion efficiency of G1 must be high if cluster gas is the source of the accretion. 

In the framework of Bondi accretion, we further derive the Eddington-scaled
Bondi accretion rate (see Table 1) and compare with other low luminosity supermassive black holes as shown in Figure 14 of Soria et al. (2006). According to the standard advection-dominated accretion flow (ADAF) model, the X-ray luminosity of G1 should be two orders of magnitude lower. This implies that the putative IMBH of G1 is extremely efficient in accreting cluster gas. It is worth noting that apart from 47 Tuc (Freire et al. 2001), we do not have evidence of cluster gas in Galactic globular clusters (van Loon et al. 2006).  Furthermore, if ionised cluster gas (as described in Pooley and Rappaport 2006) is the only source of inflow, such a high accretion rate would be difficult to reconcile with the ADAF prediction.   
It is therefore indicative that if the X-ray emission of G1 is from an IMBH, it is more likely due to accretion from a companion star. 
However, based on the current \chandra\ data, we cannot distinguish between the two possible mechanisms (IMBH or ordinary LMXB) for generating the X-ray emission of G1.

We can rule out that the X-ray emission is from combination of low luminosity sources. The total of all low luminosity sources in 47 Tuc is only $3.7\times10^{33}$\lum, implying that the encounter rate of G1 must be $> 500$ times of 47 Tuc. However, Pooley \& Rappaport (2006) estimated that the encounter rate of G1 is only 17 times of 47 Tuc. Hence the X-ray emission must come from one or a few bright sources.  
Supernova remnants may also be responsible for the X-ray emission but the probability of a core collapse supernova in a globular cluster is extremely small, as is the probability of a Type Ia supernova (e.g. Pfahl et al. 2009). Indeed, if the X-ray source is associated with the radio source detected with the VLA (Ulvestad et al. 2007), both the LMXB and supernova remnant scenarios are unlikely based on the radio/X-ray flux ratio (see below).
The probability of positional coincidence between G1 and an unrelated X-ray source is very small. We estimated the background AGN/foreground star rate within the half-mass radius of G1 ($\sim 1''$; Barmby et al. 2007) using the \chandra\ Deep Field (Brandt et al. 2001). At an 0.5--2 keV flux of $1.2\times10^{-14}$ erg s$^{-1}$ cm$^{-2}$, we expect to find on average $4\times10^{-5}$ background or foreground objects within the half mass radius.  

Using the VLA, Ulvestad et al. (2007) detected a radio
source coincident with G1 at 8.4 GHz with a flux density of $28\pm6 \mu$Jy. However, the radio position is offset from the X-ray and optical centre by about $1\farcs3$ (see Figure 1) using our refined \chandra\ and \hst\ observations. Because of the short baseline (3.5 km) of the VLA observations, the rms error of the VLA position is about $0\farcs6$ in each dimension. Therefore, the cluster core is still within the 95 per cent error radius (see Figure 1). In order to confirm if the radio source is associated with the X-ray source, we will require high-resolution and high-sensitivity radio observations using the Expanded VLA (EVLA) in its long baseline configuration, e-MERLIN, or VLBI. 

If we can confirm that the radio source is related to the X-ray source, this will provide a strong support that there is an IMBH near the centre of G1 because the radio/X-ray flux ratio of G1 ($\sim 5\times10^{-5}$) is several hundreds times higher than that of a LMXB, but is consistent with a $2\times10^4 M_\odot$ IMBH using the relationship in the ``fundamental plane'' linking radio/X-ray flux ratios and black hole masses (Merloni et al. 2003; see also Table 1). The radio/X-ray flux ratio of G1 is also substantially lower than that of supernova remnants ($\sim 10^{-2}$) and low-luminosity active galactic nuclei (see Ulvestad et al. 2007 and Table 1). 
The radio/X-ray flux ratio is consistent with a pulsar wind nebula, though we have no evidence for pulsars with such high rotational energy losses having been born in old populations such as G1.  A radio spectrum (with EVLA or e-MERLIN) or milliarcsecond-resolution VLBI observations, as suggested by Ulvestad et al. (2007), could rule this possibility out, as could detection of strong X-ray variability.
It is also possible that the radio source is not coincident with the X-ray source. In this case, the radio source may be a background source, a radio-loud X-ray binary in G1, or a jet from the central IMBH.

\section*{Acknowledgments}
This research has made use of data provided by the Chandra X-ray Center, which is operated by the Smithsonian Astrophysical Observatory on behalf of NASA, and is based on observations made with the NASA/ESA Hubble Space Telescope, and obtained from the Hubble Legacy Archive, which is a collaboration between the Space Telescope Science Institute (STScI/NASA), the Space Telescope European Coordinating Facility (ST-ECF/ESA) and the Canadian Astronomy Data Centre (CADC/NRC/CSA).
Ground-based observations were obtained with MegaPrime/MegaCam, a joint project of CFHT and CEA/DAPNIA, at the Canada-France-Hawaii Telescope (CFHT) which is operated by the National Research Council (NRC) of Canada, the Institut National des Science de l'Univers of the Centre National de la Recherche Scientifique (CNRS) of France, and the University of Hawaii.	Access to the CFHT was made possible by the Institute of Astronomy and Astrophysics, Academia Sinica, Taiwan. The CFHT data products were produced at the TERAPIX data centre located at the Institut d'Astrophysique de Paris. 
A.K.H.K. thanks L. Sjouwerman for discussion. 
This project is supported by the National Science Council of the Republic of China (Taiwan) through grant NSC96-2112-M007-037-MY3.

\end{document}